\documentclass[a4paper,11pt]{article}
\usepackage{pos}

\usepackage{etoolbox}  
\apptocmd{\thebibliography}{\setlength{\itemsep}{0pt}\setlength{\parskip}{0pt}}{}{}

\title{TeV Emission from PSR B1055-52 with H.E.S.S.: Evidence for a Pulsar Halo}

\author*[a]{Tina Wach}
\author[a]{Alison Mitchell}
\onbehalf{for the H.E.S.S. Collaboration}

\affiliation[a]{Erlangen Centre for Astorparticle Physics,\\
   Nikolaus-Fiebiger-Straße 2, 91058 Erlangen}

\emailAdd{tina.wach@fau.de}
\emailAdd{alison.mitchell@fau.de}

\abstract{Pulsar halos are a recently identified class of TeV $\gamma$-ray sources, offering valuable insights into the evolution of pulsar systems at the highest energies. However, only a handful of such sources have been detected so far, making each new identification critical for understanding the properties of the population as a whole. We report the first detection of extended very-high-energy (VHE) $\gamma$-ray emission around PSR~B1055$-$52 using observations from the H.E.S.S. array. This middle-aged pulsar, previously grouped together with Geminga and PSR~B0656$+$14 as part of the ``Three Musketeers'', has now been confirmed to host a TeV pulsar halo, making it the third detected system of its kind, and the first TeV pulsar halo discovered in the southern hemisphere. Our analysis performed in an energy range of $0.3-60\,$TeV, reveals gamma-ray emission with a one sigma extension of $(2.05 \pm 0.32)^\circ$. The analysis indicates that the emission extends beyond the region which was observed with H.E.S.S.. No significant spectral variation is detected across the emission.

The diffusion coefficient derived for this halo is significantly lower than the standard ISM value, aligning with findings in the Geminga halo and indicating that slow diffusion may be a common property of pulsar halos. The detection of this new TeV pulsar halo provides a crucial data point for studying the population-wide properties of pulsar halos, their impact on cosmic-ray propagation, and their role as a source of Galactic electrons and positrons.}

\ConferenceLogo{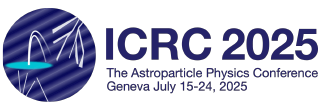}

\FullConference{39th International Cosmic Ray Conference (ICRC2025)\\
 15–24 July 2025\\
Geneva, Switzerland\\}

\begin{document}
\maketitle

\section{Introduction}

Pulsars lose rotational energy primarily through winds of relativistic particles, which interact with the surrounding medium to form a termination shock. This shock front acts as an acceleration region for subsequent relativistic particles, which produce $\gamma$-rays via synchrotron radiation or inverse Compton scattering. In young systems, this emission manifests as a pulsar wind nebula (PWN), detectable from X-ray to $\gamma$-ray energies. 
In older systems, reduced spin-down power and outward transport of relativistic electrons lead to more extended structures with a lower surface brightness, pulsar halos \citep{halo_theory}. The most prominent examples of such halos have been detected around Geminga (PSR J0633+1746) and Monogem (PSR B0656+14) \citep{geminga_detection}, by the Milagro Observatory and the High Altitude Water Cherenkov Observatory. Both instruments are wide-field Water Cherenkov Detectors (WCDs) capable of surveying large sky regions continuously. 

A third pulsar, PSR B1055$-$52, was discovered in 1972 via radio observations by the Molonglo Observatory \citep{detection_b1055}, and later detected as $\gamma$-ray pulsar by the Compton Gamma-Ray Observatory \citep{detection_b1055_gamma}. With a spin-down power of $\dot{E} = 3.0 \cdot 10^{34}\,$erg/s, a surface magnetic field strength of $B = 1.1 \cdot 10^{12}\,$G, and a characteristic age of $\tau_c = 535\,$kyr \citep{atnf}, it closely resembles Geminga and Monogem in its physical properties and has, therefore, long been anticipated to host a TeV pulsar halo/

However, its sky position ($\text{R.A.} = 164.50^\circ$, $\text{Dec.} = -52.45^\circ$ \citep{atnf}) renders it inaccessible to currently active WCD arrays, making it only observable in $\gamma$-rays from the southern hemisphere by the High Energy Stereoscopic System (H.E.S.S.), an Imaging Atmospheric Cherenkov Telescope (IACT) array located in Namibia.  IACTs have in the past struggled to detect sources with extension comparable to the known TeV halos due to their limited fields of view and the extended nature of the emission. 

In this study, a specialized background estimation technique tailored for extended source analysis was applied to archival H.E.S.S. data of the PSR B1055$-$52 region. Due to these improvements, this study can report the discovery of extended TeV $\gamma$-ray emission in the region. Furthermore, we estimate the diffusion parameters of the relativistic electrons necessary to explain the observed extension of the $\gamma$-ray emission. 

\section{Data selection and Background estimation}

H.E.S.S. is an array of five IACTs located at an altitude of $1800\,$m, in the Khomas Highlands in Namibia. 
This study uses data from January to March 2019 acquired with the four $12\,$m telescopes of the H.E.S.S. array. 
The available observations, conducted in $~28\,$minute observation runs, were filtered following the quality recommendations in \citep{crab_2006}, such that only the runs acquired under good atmospheric conditions and a good system response remain for the analysis. After this quality selection $44.9\,$hrs of deadtime corrected observation time remain.  

The gamma-hadron separation was performed using a method described in \citep{gamma-hadron} and reconstructed using the Image Pixel-wise fit for Atmospheric Cherenkov Telescope algorithm \texttt{ImPACT} (for more information see \citep{impact}). Using the python package \texttt{Gammapy, version 1.2} \citep{gammapy, gammapy_zenodo}, a 3-dimensional likelihood analysis of the data is carried out. 

\subsection{Background estimation}

A significant challenge in detecting large, extended structures with IACTs is the estimation of residual cosmic-ray background since most conventional background estimation relies on OFF regions presumed free of $\gamma$-ray emission. 
This is problematic for extended sources, where assumptions about the source morphology can introduce bias.

A more robust approach is the use of a 3D background model template. The background template model used in this study is constructed from archival H.E.S.S. data (for more information, see \citep{bkg_template}) and adjusted for the respective observation conditions via a fit of a flux normalization $\Phi$ and spectral index $\delta$, to $\gamma$-ray-free regions:
\begin{equation}
    R^*_{\mathrm{BG}} = \Phi \cdot R_{\mathrm{BG}} \cdot \left(\frac{E}{E_0}\right)^{-\delta}\,.
\end{equation}

While robust, this background estimation still requires a region without $\gamma$-ray contamination for scaling. If no such region can be found, the background will be overestimated, consequently leading to an underestimation of total source flux and morphology. 

To alleviate this disadvantage, the background can be scaled not on a $\gamma$-ray-free region within the observation, but on an OFF run, a separate observation acquired under similar observation conditions and with only minimal, well-known $\gamma$-ray contamination. The background rate is then directly inferred from the OFF run, eliminating the need for a source-free region in the target run's field of view. Thanks to large number of archival observations which were used to create the background model template, this approach is more stable than a traditional ON/OFF background estimation. Details of this method and the matching criteria are provided in \citep{rm}.




\section{Analysis Results}

\begin{figure*}
\begin{minipage}[b]{.49\textwidth}
\includegraphics[width=0.99\textwidth]{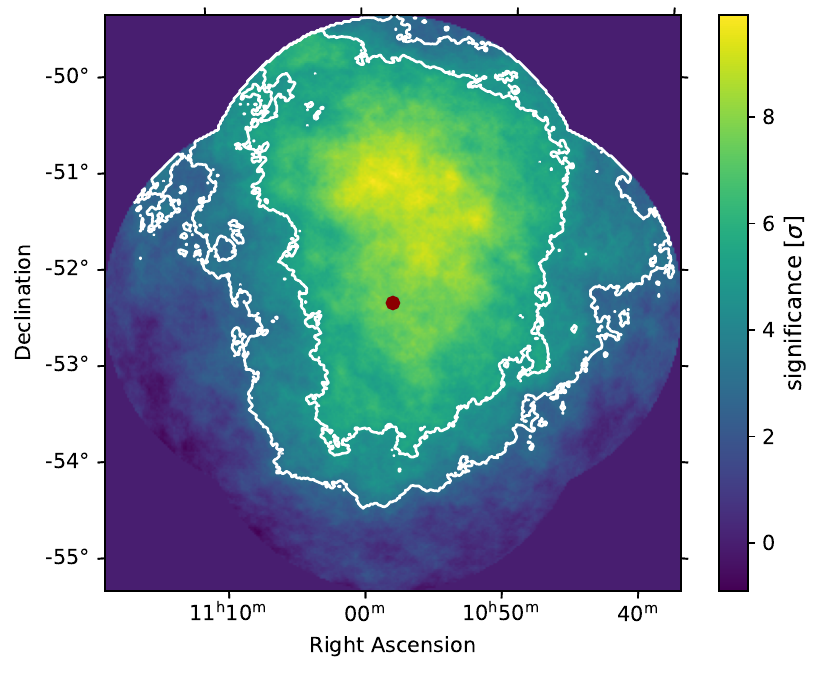}
\end{minipage}\qquad
\begin{minipage}[b]{.49\textwidth}
\includegraphics[width=0.99\textwidth]{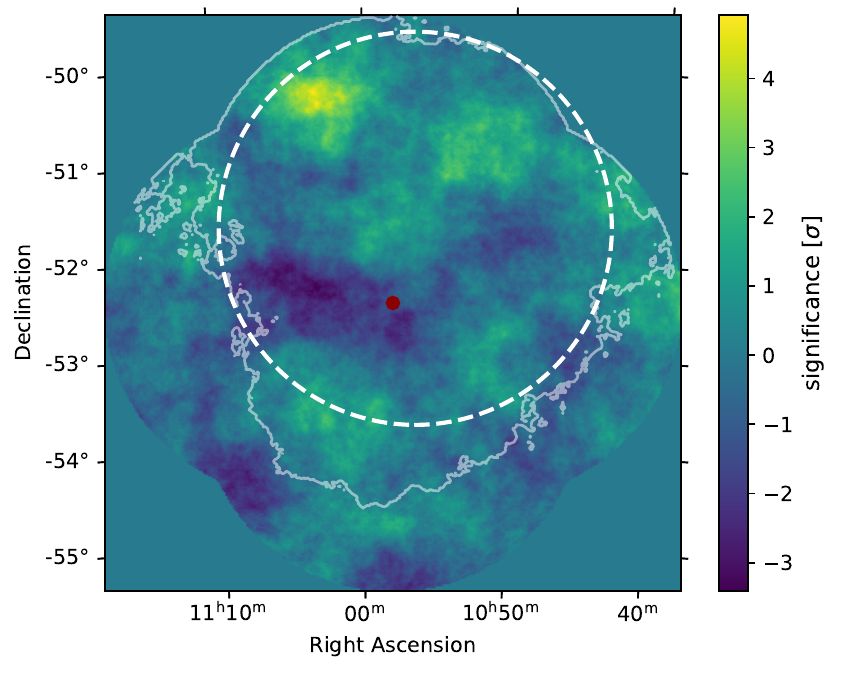}
\end{minipage}
\caption{Significance map of the region around PSR~B1055$-$52 with a $1\deg$ correlation radius. Left: The $3\,\sigma$ and $5\,\sigma$ contours of the emission are shown in white, while the pulsar position is indicated in red. Right: Residual significance map after the model fit. The best-fit morphology of the Gaussian model is indicated by the dashed circle.}
\label{fig:threshold_choice}%
\end{figure*}

Figure \ref{fig:threshold_choice} shows a Li\&Ma significance map computed with a correlation radius of $1\,^\circ$. On the left, the $3\,\sigma$ and $5\,\sigma$ contours of the emission are shown in white. 
The significance map illustrates that the emission is not fully contained within the observed region, and many of the results presented in this study, especially the source extension, should be interpreted as a lower limit. 

\begin{table}
\caption{Best-fit parameters obtained for the analysis of the data above the energy threshold, using a Gaussian model with a power law as spectral model. The centre of the Gaussian model was fixed to the centre of gravity of the $\gamma$-ray emission in the region for this fit.} 
\label{tab:fitparamters} 
\centering                          
\begin{tabular}{c | c }       
\hline\hline  
\noalign{\smallskip}
 & PSR~B1055$-$52 region  \\    
\noalign{\smallskip}
\hline 
\noalign{\smallskip}
   $\Gamma$  & $2.80 \pm  0.12_\text{stat} \pm 0.42_\text{sys}$   \\[0.1cm]
$\text{N}_0\,\,[10^{-11} \text{cm}^{-2}\,\text{s}^{-1}\,\text{TeV}^{-1}]$  & {$2.72 \pm 0.40_\text{stat} \pm 1.57_\text{sys}$}   \\[0.1cm]
   R.A. $[^\circ]$ & $164.12 \pm 0.35_\text{sys}$  \\[0.1cm]
   Dec. $[^\circ]$ & $-51.67 \pm 1.40_\text{sys}$  \\[0.1cm]
   $\sigma$ $[^\circ]$ & $2.05 \pm 0.32_\text{stat} \pm 0.72_\text{sys}$  \\[0.1cm]
\noalign{\smallskip}
\hline   
\hline 
\end{tabular}
\end{table}

Due to the emission not being fully contained, no spectromorphological fit of a source model to the data is possible. To circumvent this problem, the centre of gravity of the $\gamma$-ray excess is computed and used as the centre of a Gaussian source model. Then the extension of this Gaussian, as well as the spectral parameters of a power-law model are fitted to the data. The best description of the $\gamma$-ray emission around PSR~B1055$-$52 can be reached by applying a symmetric Gaussian model. The best-fit parameters can be seen in Table \ref{tab:fitparamters}.  A significance map of the emission, overlaid with the best-fit morphology of the model, and the $3\,\sigma$ contours of the emission can be seen in the right panel of Figure \ref{fig:threshold_choice}. 

The significance map obtained after fitting the source model (right panel of Figure \ref{fig:threshold_choice}) shows a slight remaining excess north of the pulsar, indicating residual significant emission. An additional Gaussian component was introduced to account for this excess, but it was not found to be statistically significant.



\subsection{Systematic uncertainty on the detection significance}

Several tests were performed to show the robustness of the detection and estimate the uncertainty on the source detection and best-fit parameters.

To approximate the systematic uncertainties on the background estimation, two additional datasets were created. For the first dataset, different ON-OFF pairs were selected to estimate the uncertainty introduced by the choice of run pairs. For the second dataset, a Gaussian prior of $5\%$ was added to the normalization of the background model template. Then, the Gaussian source model was fitted to the data, and a source significance was derived for both cases. We find that the significance of the source marginally decreases to $15.37\,\sigma$ for the run list with different run pairs and marginally increases to $15.97\,\sigma$ when the Gaussian prior is included. This indicates a stable background estimation. 



A conservative uncertainty on the source parameters (shown in Table \ref{tab:fitparamters}) was estimated from the fit of the Gaussian source model to a dataset for which the background was estimated through a direct application of the background model template. Due to the lack of $\gamma$-ray free regions, the background rate, in this case, is overestimated, but this result can still be used as a conservative lower limit on the extension and flux of the emission.

\subsection{Energy dependence of the emission}

To investigate a possible energy dependence of the extension of the $\gamma$-ray emission, the data were divided into four logarithmically spaced energy bands and the Gaussian source model fitted in each band. In the first two energy bands, the centre of the model was fixed to the centre of gravity of the emission calculated in the respective energy band. For the other two energy bands, the position of the model was included as a fit parameter. 

\begin{figure}
\centering
\includegraphics[width=0.5\textwidth]{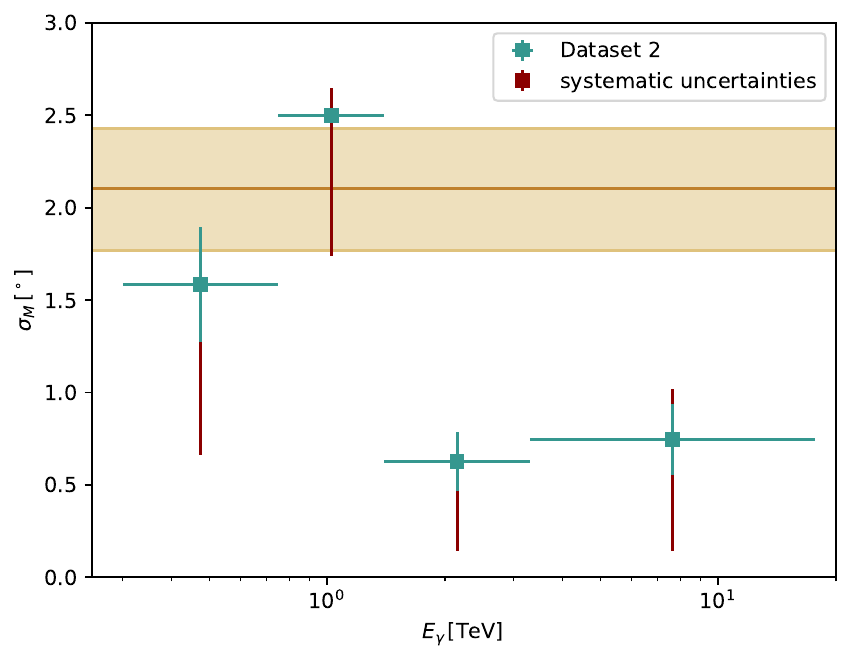}
\caption{Extension of the Gaussian model fitted in the different energy bands.}
\label{fig:enedep}
\end{figure}

Figure \ref{fig:enedep} compares the $1\,\sigma$ containment radius for the Gaussian model for all energy bands. The best-fit extension derived from a fit over the whole energy range is shown by the beige band. 

The spatial extension of the emission exhibits a peak near $1\,$TeV, which is likely an artifact introduced by fitting emission that extends beyond the observed region. At energies above $1\,$TeV, a decrease in the extension is observed. However, due to the limitations of this dataset, no conclusive trend with energy can be observed. 


\section{Interpretation}

The region around PSR~B1055$-$52 observed by H.E.S.S. includes two pulsars, PSR~B1055$-$52 and PSR~J1103$-$5403, located at R.A. = $165.8875\,^\circ$, Dec = $-54.0619\,^\circ$. The latter is an old millisecond pulsar with a characteristic age of $\tau_c = 1.46 \times 10^7\,$kyr, a spin-down luminosity of $\dot{E} = 3.7 \times 10^{33}\,$erg/s and a distance of $1.68\,$kpc, which is likely insufficient to power significant VHE $\gamma$-ray emission. The region also contains several molecular clouds \citep{clouds3, clouds1}. However, none of these molecular clouds can account for the full spatial extent of the observed $\gamma$-ray emission. It is therefore plausible to assume that the origin of the $\gamma$-ray emission is relativistic electrons originating from PSR~B1055$-$52.

Due to the decay of the magnetic field strength over the age of the pulsar, a distinction between pulsar wind nebula and pulsar halo can be made. A young pulsar is expected to dominate its surrounding environment, leading to a confinement of leptons in a highly magnetized bubble around the pulsar and an energy density of leptons higher than the energy density expected in the ISM. For older pulsars, it is generally expected that the pulsar itself no longer dominates the surrounding medium. The leptons can freely diffuse outward and cool, leading to an energy density comparable or lower than the energy density of the ISM $\epsilon_e \lesssim 0.1\,\text{eV/cm}^3$. 

The energy density in the region around the pulsar can be estimated using the spin-down power, the characteristic age and the extension of the region containing the $\gamma$-ray emission as $\epsilon_e  =  \frac{\dot{\text{E}}\tau_c}{4/3\pi R^3}$ with $R$ the $3\,\sigma$ containment radius of the Gaussian, $3\sigma_M$, and the distance $d$ to the source as $R = 3\sigma_M \cdot d$. For this study, a distance estimate of $d=350\,$pc from \citep{distance_b1055_new} is adopted. The energy density for the emission at $350\,$pc is derived to  $\epsilon_\text{e, 350} = (4.84 \pm 1.68) \cdot 10^{-2}\,\text{eV/cm}^3$, which is is significantly below the typical ISM value. It is therefore likely that the region is not dominated by strong pulsar magnetic fields, and the leptons undergo diffusive transport away from the pulsar.

Despite this low energy density and the implied consequence that the region is not dominated by the magnetic field of the pulsar, studies have shown that the diffusion in the region around the Geminga and Monogem pulsars are suppressed with respect to the ISM \citep{hawc_geminga, geminga}, indicating that the pulsar still influences its surroundings. To test whether this is also the case for the region around PSR~B1055$-$52, a diffusion coefficient is calculated from the observed extension of the $\gamma$-ray emission in energy bands, and the cooling time of the electrons using $r_\text{diff} = \sqrt{2 t_\text{cool} \cdot D}$, with $D$ the diffusion coefficient and $t_\text{cool}$ the cooling time of the leptons. 
The cooling time was estimated as:
\begin{equation}
    t_{\text{cool}} = \frac{m_e c^2}{\frac{4}{3} c \sigma_T \gamma} \cdot \frac{1}{u_B + u_{\text{ph}} / (1 + 4 \gamma \varepsilon_0)^{3/2}}
\end{equation}
with $u_\text{ph}$ the energy density of the photon fields, $\varepsilon_0$ the normalized photon energy and $u_B = B^2/8\pi$ with $B$ the magnetic field. Since the emission extends beyond the observed region for lower energies, only the extension derived in the last two energy bands is used for the calculation of $D$. The diffusion coefficient at the respective electron energy of the bands $\text{E}_e$ was then scaled to $100\,$TeV, assuming Kolmogorov diffusion, to allow for a direct comparison with the results derived for the pulsar halo around the Geminga pulsar, as well as the ISM diffusion coefficient of $4.5 \cdot 10^{29}\,\text{cm}^2/\text{s}$ \citep{ISM_diff}. 

In this study, the average line-of-sight magnetic field at the pulsar position was estimated using the relation:
\begin{equation}
    B = 1.23 \cdot \frac{RM}{DM}
\end{equation}
with $RM$ the rotation measure at the pulsar position and $DM$ the dispersion measure \cite{rotation_to_mag}. The rotation measure was extracted from the Galactic Faraday depth sky published in \citep{faraday}, while the $DM$ was estimated using two widely used electron density models, \citep{electron_density1} (NE2001) and \citep{electron_density2} (YMW16). The resulting magnetic field estimates and corresponding diffusion coefficients are summarized in Table~\ref{tab:diffusion_coeffs}.

\begin{table*}
\caption{Magnetic field and Diffusion coefficient at an electron energy of $100\,$TeV assuming Kolmogorov diffusion.} 
\label{tab:diffusion_coeffs} 
\centering                          
\begin{tabular}{ c | c c  }       
\hline\hline  
\noalign{\smallskip}
  & $B$ 
  & $D_0(\delta = 0.33)$  \\ 
  &  $[\mu G]$ 
  & $[10^{27}\,\text{cm}^2/\text{s}]$  \\ 
\noalign{\smallskip}
\hline 
\noalign{\smallskip}
    YMW16  & $0.64$ & $0.32 \pm 0.12$ \\[0.1cm]
    NE2001 & $8.48$ & $1.28 \pm 0.50$  \\[0.1cm]
\noalign{\smallskip}
\hline   
\hline 
\end{tabular}
\end{table*}

The diffusion coefficient for both electron density models shows significant suppression compared to the ISM diffusion. In particular, for the YMW16 model \citep{electron_density1}, the derived coefficient lies below the Bohm limit, which corresponds to the slowest possible diffusion consistent with quasi-linear theory. In order not to violate the Bohm limit, a magnetic field strength of $B > 5.1\,\mu$G would be required, assuming standard particle transport and the absence of extreme turbulences. Alternatively, for the field inferred in \citep{electron_density1} to be consistent with transport, the pulsar would need to lie at a much larger distance ($>3.4\,$kpc).

However, it is important to note that both $RM$ and $DM$ measurements yield only an electron-density-weighted average of the magnetic field along the line of sight, and are insensitive to components perpendicular to the line of sight. If the true magnetic field is predominantly transverse, then the inferred field may substantially underestimate the total field strength. Moreover, field reversals or local fluctuations along the line of sight could further suppress $RM/DM$, complicating interpretation.

\section{Conclusion and Outlook}

This work details the discovery of extended emission around PSR~B1055$-$52 with H.E.S.S., employing a novel background estimation technique optimized for the detection of large $\gamma$-ray structures. Due to insufficient coverage of the region, the full extent of the source remains unresolved, but a minimal $39\%$ containment radius of $\sigma = 2.05 \pm 0.32^\circ$ is measured. Notably, the centroid is offset by $\sim 5\,$pc from the pulsar position (assuming a distance of $350\,$pc), in a direction nearly perpendicular to the pulsar’s estimated proper motion. This may suggest anisotropies in electron escape or propagation.

We also observe indications of energy-dependent morphology, consistent with diffusion and cooling of electrons injected by a point-like source. While the observed energy range and angular extent are limited, this trend aligns with findings in other evolved systems around pulsars such as Geminga and HESS~J1825$-$137.

The detection of extended TeV emission around PSR~B1055$-$52 supports long-standing expectations that this middle-aged pulsar hosts a pulsar halo \citep{1055_expect}.  
Adding PSR~B1055$-$52 to the class of pulsar halos will help to refine our understanding 
and make more meaningful comparisons across the population. One such property is the finding of inhibited diffusion around the Geminga halo. This work finds that inhibited diffusion can also be observed around PSR~B1055$-$52, potentially having major implications for cosmic-ray propagation.

To fully characterize this source, a diffusion model will be employed to derive the properties of the parent electron population necessary to produce the observed $\gamma$-ray emission. In addition to that, future wide-field observations of the southern sky are essential. Upcoming instruments such as ALPACA and SWGO will be crucial for resolving the full morphology and refining the diffusion parameters.

\begingroup
\footnotesize
\section*{Acknowledgments}
This work is supported by the Deutsche Forschungsgemeinschaft (DFG, German Research
Foundation) – Project Number 452934793 and was co-funded by a program supporting faculty-specific gender equality targets at Friedrich-Alexander University Erlangen-Nürnberg (FAU).
The full acknowledgments of the H.E.S.S. Collaboration can be found at \url{https://hess.in2p3.fr/acknowledgements/}. 

\bibliographystyle{JHEP}
\bibliography{mybibliography}
\endgroup

\end{document}